\documentstyle[multicol,aps,epsf]{revtex}   
\begin{document}
\title{Disorder driven destruction of a phase transition in a superconductor}

\draft
\author{N.K. Wilkin \cite{add} and Henrik Jeldtoft Jensen}

\address{Department of Mathematics,\\ Imperial College, 180 Queen's Gate,\\
London, SW7 2BZ, United Kingdom}

\date{\today}
\maketitle
\begin{abstract}
We investigate the effects of disorder on a layered
superconductor. The clean system is known to have a first order phase
transition which is clearly identified by a sharp peak in the specific
heat. The peak is lost abruptly as the strength of the disorder is
increased. Hence, for strong disorder there is no phase transition as
a function of temperature but merely a crossover which is still
detectable in the IV characteristic.
\end{abstract}
\begin{multicols}{2}
Disorder is known to play an important role in the phase diagram of
even so-called {\em clean} superconductors at low temperatures and
high magnetic fields. It has recently been show that these clean
systems have a first order transition. It is believed that this
transition is associated with melting or decoupling of the vortex
system. The thermodynamic signal of the transition is lost at a
`critical point' below which pinning is thought to dominate the
behaviour\cite{Zeldov_Nat_95,Khaykovich_96}. 

Significant effort is currently being invested in attempting to
understand theoretically the effect of disorder on the behaviour of
the magnetic flux system in
superconductors\cite{JaglBals97,Kier97}. In this letter we discuss
simulations of the phenomenological behaviour and response to disorder
of a layered vortex system. Our model is deliberately made
sufficiently simple that we are able to  identify the
mechanisms behind the effects we observe. We believe that our results
demonstrate which degrees of freedom of the vortex system are
necessary in order to interpret the real experimental data.

We present the results of a 3D layered simulation in the presence of
point disorder.  We have included {\em only} the degrees of freedom
associated with vortex lines, an approximation we believe to be valid
away from the vicinity of the zero field transition, $T_c$. The vortex
lines consist of stacks of pancake vortices.  These stacks are able to
cross and to decouple when it becomes energetically favourable and
this is an essential feature of the model. The clean system has a
first order decoupling transition, with an entropy jump comparable to
that seen experimentally away from $T_c$: namely $\simeq 0.4
k_B$/pancake/layer. However, for strong disorder the first order
transition is reduced to a gradual crossover.  This can be seen in
many aspects of the behaviour. The pronounced peak in the specific
heat that is a feature of the clean system is no longer present and
the diffusion becomes thermally activated with an energy scale set by
the strength of the interlayer coupling. Thus the decoupling
transition is replaced by a depinning crossover whose underlying
mechanism is plastic cutting of the vortex lines. This is confirmed by
direct measurement of the cutting frequency. Hence a {\em vortex
glass} phase \cite{FFH} {\em cannot} occur in this system since the
divergence of the elastic creep barriers at vanishing driving
force\cite{Feig89} will be cut-off by the finite activation energy
barrier for plastic cutting of the flux lines. We analyze both
thermodynamic and transport properties. The specific heat behaviour
demonstrates that above a threshold disorder destroys the phase
transition of the clean system. However, the transport properties are
relatively insensitive to the disorder. This stresses that one must be
cautious when inferring the existence of phase transitions from
transport data \cite{Jag96}. The manner in which the clean system phase
transition is destroyed by the introduction of disorder has been
investigated by Kierfeld {\em et al.}\ \cite{Kier97,Kier97b}. The destruction
of the thermodynamic phase transition may in fact take place via a
phase transition as a function of disorder in our simulation.

The model is a layered pancake system which demonstrates quantitative
features similar to those seen in the highly anisotropic Bi-2212 high
temperature superconductor. The transition is studied as a function of
fixed magnetic field, B (density of vortices), which is always
perpendicular to the layers (along the c-axis), and variable
temperature. In order to be able to study the loss of the vortex
lattice order the vortex position are varied continuously as an
underlying discrete lattice can lead to spurious
phases\cite{Hattel_95}.  The disorder is the simplest possible ---
namely, random point pins.

The system is equilibriated via Langevin dynamics with periodic
boundary conditions enforced in all directions. For simplicity all the
temperature dependence of the model is introduced via a noise term
\cite{Brass_89}; additional temperature dependencies of the
penetration depth and other length scales are neglected. We focus
purely on the vortex lattice aspect of the melting -- that is only
vortex loops representing fluctuations in the positions of the flux
lines are included. To enable a sufficiently large simulation to study
3D effects Gaussian potentials are employed for all the
interactions. In 2D the true in-plane interaction should be the $K_0$
Bessel function but it is only reasonable to use this numerically for
very short penetration depths as otherwise the long range nature of
the potential results in excessive relaxation times. For a layered
superconductor the predicted potentials \cite{Blatter_94} are even
longer. That the qualitative behaviour is correctly simulated by the
Gaussian potentials can be seen by comparing the 2D simulations of
Jensen {\em et al.}\ \cite{Jensen_88} who used the Gaussian potential and
Koshelev \cite{Koshelev_92} who used the physical potentials.

The pancakes in the planes have an in-plane repulsive interaction
between them, in our case modelled by a Gaussian potential $U_{{\rm
v}{\rm v}'}^{ll}=A_{{\rm v}}\exp(-r^2/\xi_{\rm v}^2)$, where $r$ is
the in-plane distance between the vortices, $
\xi_{{\rm v}}$ is the in-plane vortex range, and $A_{\rm v}$ is the
(fixed) strength of the vortex potential.
 
Across the layers the interaction is more complicated as it must
include a mechanism for cutting and reconnecting the vortices. This is
known to be important near the melting temperature, in order to allow
for the loss of long range phase coherence as seen, for example  in the
pseudo-transformer experiments
\cite{delacruz_94}. It has been shown by Clem
\cite{Clem_91} that in order to model the electromagnetic interactions
across the layers only pair-wise potentials are needed. However,
Bulaevskii {\em et al.}\ \cite{Bulaevskii_92a} have shown that
including the lowest order terms of the Josephson coupling makes three
and four-body terms equally necessary. For weak interlayer coupling we
have found that the inclusion of the three body term is
sufficient. Hence the total interaction across the planes is composed
of a two body attractive interaction (where all $r$ are in-plane)
$U_{{\rm v} {\rm v}'}^{l l'}(r_i^l,r_j^{l'})= -A_{\rm
l}\exp(-(r_i^l-r_j^{l'})^2/\xi_{l^2}) $ and a three body repulsive
interaction.
\begin{equation}
U_{{\rm v} {\rm v}'{\rm v}''}^{l l l'}(r_i^l,r_j^l,r_k^{l'})= A_{{\rm 3b}} 
e^ {- \left((r_i^l-r_j^l)^2+(r_j^l-r_k^{l'})^2
+(r_k^{l'}-r_i^l)^2\right)/\xi_{\rm 3b}^2}.
\end{equation}

$A_l$ and $\xi_l$ are the amplitude and range of the two body
interaction and similarly $A_{{\rm 3b}}$ and $\xi_{\rm 3b}$ are the
amplitude and range of the three body potential. The latter acts by
excluding three or more pancakes (two in one layer and a third in an
adjacent layer) from finding their equilibrium location to be within a
coherence length in the x-y plane. The anisotropy is determined by the
strength of the interlayer coupling parameter $A_l$, and is fixed to
be $0.2$, which corresponds to a highly anisotropic system. That is
the ratio between the tilt and shear moduli is $c_{44}/c_{66} \simeq
0.01$ for layer and vortex spacings pertinent to Bi-2212 in an
external magnetic field of 1T. This is similar to that estimated
for Bi-2212 using the elastic modulii as defined in Blatter {\em et al.}\
\cite{Blatter_94}. The other parameters are: $A_{\rm v}=1$ and
$A_{\rm 3b}=A_l$, for the amplitudes of the potentials and $\xi_{{\rm
v}}=0.6$, $\xi_{l}=0.3$, $\xi_{{\rm 3b}}=\sqrt{2} \xi_{l}$ for the
ranges in the Gaussians.  Our unit of length is the average spacing
between the pancakes.

For this choice of parameters the clean system loses order in both the
$a-b$ plane and the c-direction over a very narrow temperature range
which is comparable to the accuracy with which the temperature can be
determined in the Langevin simulation.  There is a pronounced peak in
the specific heat at this temperature and a Lee-Kosterlitz
\cite{LeeKost91} binning of the energies yields an associated entropy
of $\simeq 0.4 k_B$/pancake. Furthermore the activation energy between
the ordered and disordered states grows with system size, indicating a
first order transition. 

In two dimensions it has been shown by one of the authors
\cite{Jensen_89} that weak disorder with amplitude in the range $0.01
\leq A_p \leq 0.05$ has a significant effect on the
melting temperature (as defined by the onset of diffusion),
Fig.~\ref{fig:diswk}. $T_{\rm ab}$ initially decreases (to
approximately half its clean value) but then increases as the pinning
energy become dominant, lower inset to Fig.~\ref{fig:diswk}. This is
due to the weakening by disorder of the shear modulus of the system.
In three dimensions this effect is still present but much weaker (of
the order of a percent), see upper inset to Fig.~\ref{fig:diswk}. At
these small values of $A_p$ there is no difference in the results
obtained for a broad range of vortex-to-pin ratios: $0.5\leq N_v/N_p
\leq 2$.

Experimentally the transition moves to lower temperatures as the
magnetic field is increased \cite{Zeldov_Nat_95}. At these lower
temperatures less thermal energy is available and hence the pinning
energies become more significant.  To simulate this without changing
the magnetic field a larger value of the pinning potential amplitude,
$A_p=0.5$ is used. The results are dramatically different from the
clean system. Firstly, the onset of the diffusion which in the clean
system is very sharp becomes a gradual onset, and is in fact thermally
activated. The activation energy is determined by the strength of the
interlayer coupling: $E \simeq 1.5
A_l$ (from investigating $A_l=0.2$,0.3 and 0.5 with fixed $A_p=0.5$
and varying numbers of pins). This suggests that the diffusion is
occurring via {\em plastic} cutting of the vortex lines. The cutting
rate can also be measured directly. We define the rate  to be the the number
of times a pancake cuts and reconnects to pancakes in adjacent
layers, in a given time interval. There is a plateau in the rate above
the pure system decoupling temperature, see Fig.~\ref{fig:cut} for all pinning
strengths. However, the onset temperature for cutting decreases as the
pinning strength increases, in agreement with the diffusion results.

In the clean system there is a peak in the specific heat, which is a
thermodynamic indicator of the transition and can be compared with the
experimental results of Zeldov {\em et al.}\ and Schilling {\em et
al.}. In the presence of weak disorder ($A_p \lesssim 0.1$) the peak
is essentially unaffected but the peak vanishes rapidly with
increasing disorder when $A_p \simeq 0.2$ and drops away into the
background, Fig.~\ref{fig:specdis}. It should be emphasized that when
a peak is present, it is as sharp as our temperature resolution. The
height of the peak is difficult to determine due to the strong
fluctuations in the vicinity of the transition. Our results are fully
consistent with the ideas of Kierfeld {\em et al.}\ \cite{Kier97} that
there may be a phase transition as a function of disorder.

We have shown that the thermodynamic phase transition is destroyed by
directly investigating a thermodynamic property (the specific
heat). We now demonstrate that the existence of the transition can be
falsely assumed by indirect measurements, such as IV characteristics.

For $A_p \gtrsim 0.2$ there is no longer a true phase transition but
this is not apparent if a diffused distance criteria, say
$R^2(t=t_s)=1$ is used to determine melting. From this criteria one
finds that disorder merely shifts the melting temperature and one
would misleadingly conclude that the melting transition survives the
presence of disorder.  It is, however, clear from the form of the
diffusion curves that the diffusion mechanism has changed, see Fig.~\ref{fig:difdis}. Measuring
the onset of diffusion is a close analogue of an experimental IV
characteristic. This is demonstrated in Fig.~\ref{fig:iv} with {\em
log-log} plots for both the diffused distance as a function of
temperature (at fixed time) and the voltage as a function of
temperature (at fixed driving force). Within the numerical accuracy
the IV characteristic is linear as a function of current at small current. The IVs 
correspond to thermally assisted flux flow (TAFF)\cite{Blatter_94}. By assuming that
the induced voltage has an Arrhenius form we find that the activation
energy is of the same order of magnitude as that deduced from the
onset of the diffusion data. Hence our system does not  have a
vortex glass phase \cite{FFH}.

To summarize we have presented a simple model which demonstrates that
the thermodynamic decoupling transition is abruptly destroyed by a finite
amount of disorder. We believe that this is the explanation for the
critical point in the Bi-2212 phase diagram of Zeldov
{\em et al.}. Furthermore, our study shows that transport data can
suggest the existence of a transition even when it is
known that disorder has destroyed the thermodynamic phase transition.

N.K.W thanks the School of Physics, Birmingham University for
hospitality whilst some of this work was carried out. N.K.W.\ and
H.J.J.\ were supported by the EPSRC grant no.\ Gr/J 36952 and GR/L28784.

\begin{figure}
\narrowtext
\caption{Main figure: Diffusion as a function of temperature for the
two dimensional system with $N_v=1024$ for different pinning
strengths, $A_p$. The insets are the 2D and 3D behaviours of the
melting temperature (as defined by diffusion) as a function of pinning
strength. It is clear that the effect is much more significant in the
2D case.}
\label{fig:diswk}
\end{figure}

\begin{figure}
\narrowtext
\caption{Number of cutting events/pancake as a function of the strength of the disorder.}
\label{fig:cut}
\end{figure}

\begin{figure}
\narrowtext
\caption{Specific heat as function of temperature for the clean  (solid
line) and with $144$ pins of amplitude $A_p=0.5$ and
range $R_p=0.125$ (dashed line). Both systems consist of
144 pancakes in each of 8 layers.}
\label{fig:specdis}
\end{figure}

\begin{figure}
\narrowtext
\caption{Average of square of moved distance over a fixed number of
time-steps for different number of pins $N_p$.The amplitude is $A_p=0.5$ and
range is $R_p=0.125$ in all cases. The system consists of
144 pancakes in each of 8 layers. The inset shows the temperature at 
which $R^2=1$ for different pin densities }
\label{fig:difdis}
\end{figure}

\begin{figure}
\narrowtext
\caption{Plots of the Log(Voltage) and Log(diffused distance) versus Log(Temperature) for $A_l=0.2$ and $A_p=0.5$ indicating an onset temperature for 
both $\sim0.04$. The insert shows the change in the behaviour of the diffusion curve between the clean and strongly pinned systems. } 
\label{fig:iv}
\end{figure}

\end{multicols}

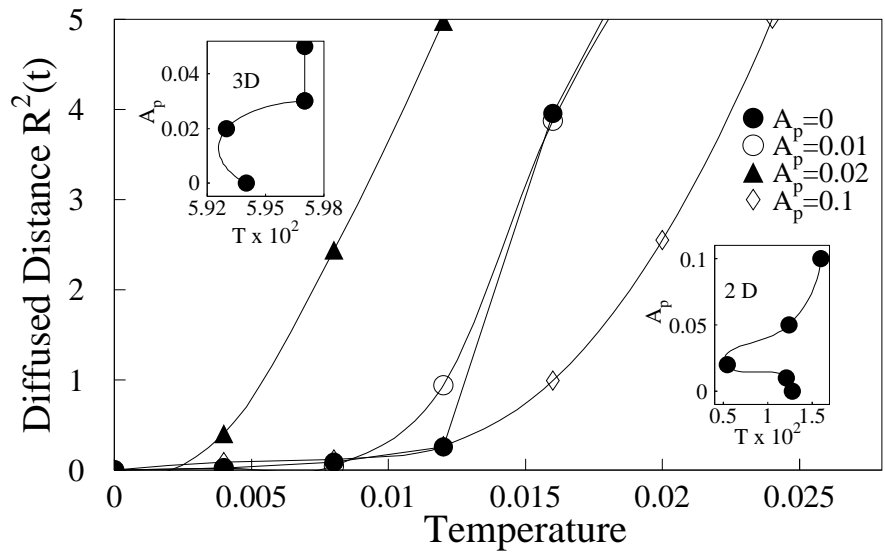
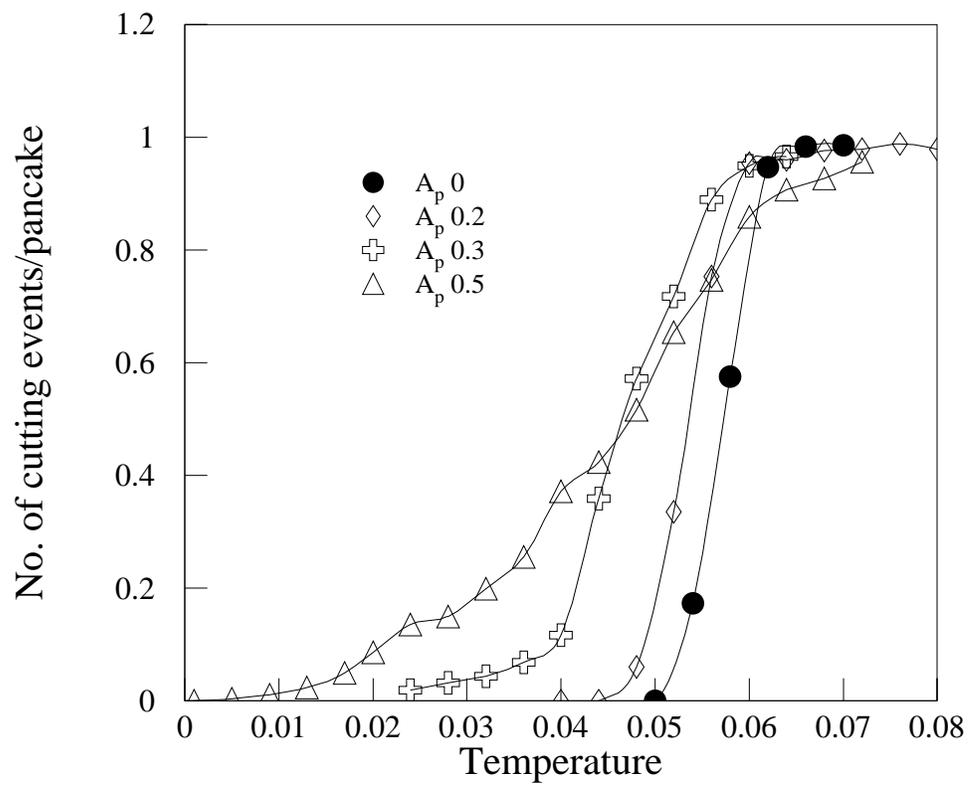
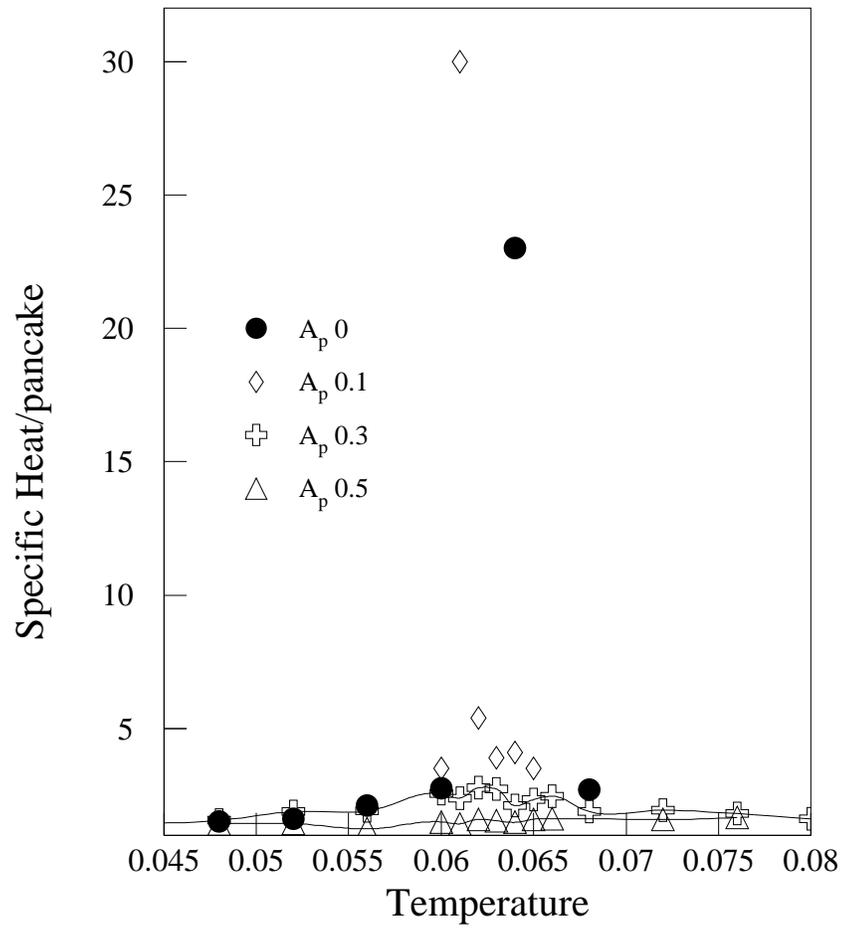
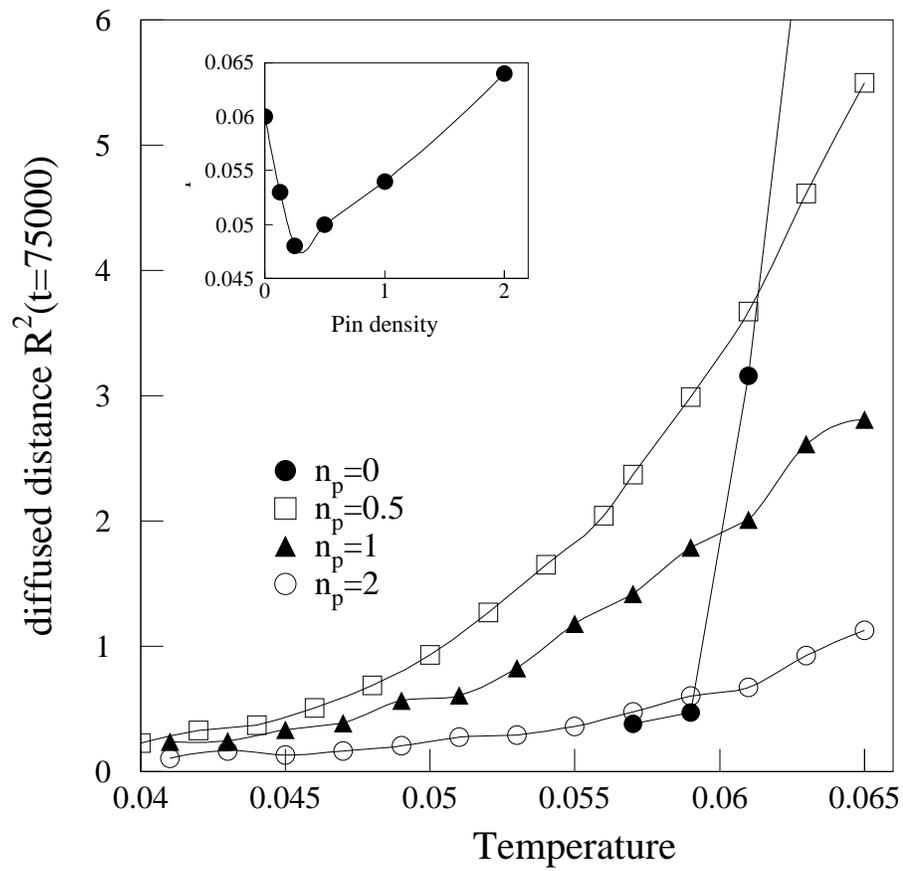
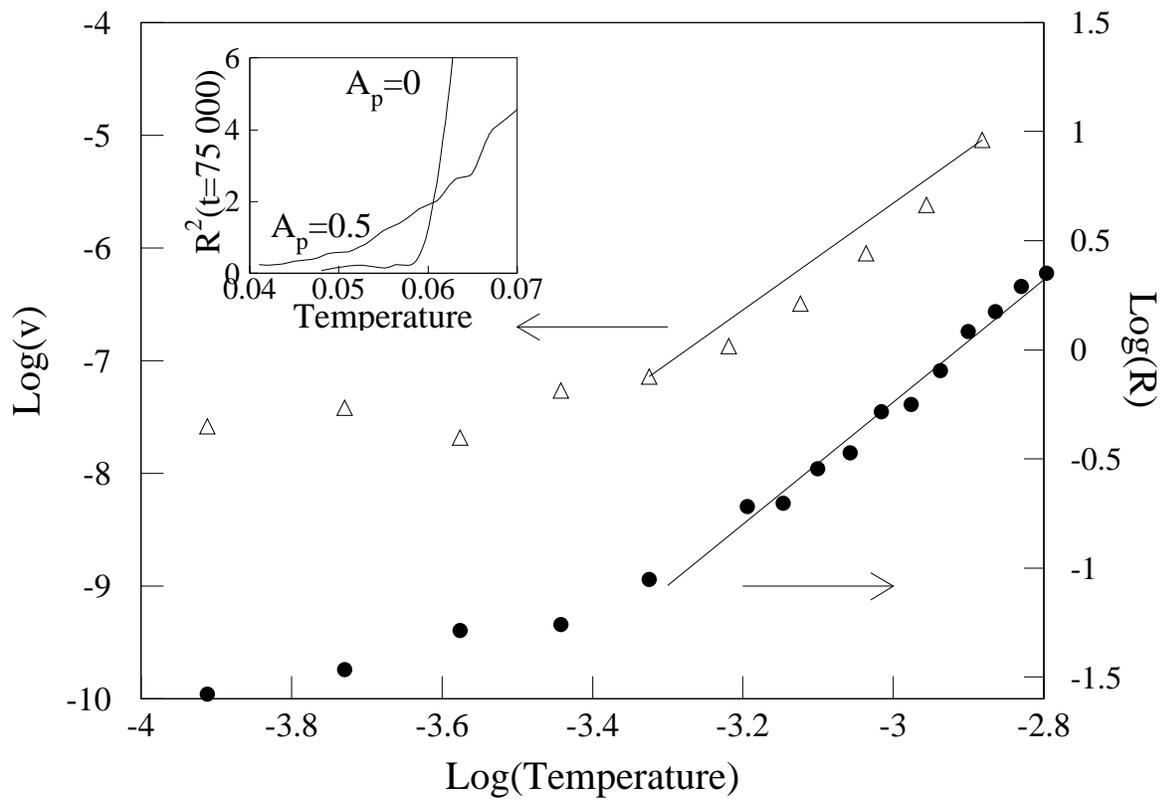
\end{document}